\documentstyle[12pt]{article}
\def\Re{{\cal R \mskip-4mu \lower.1ex \hbox{\it e}\,}}
\def\Im{{\cal I \mskip-5mu \lower.1ex \hbox{\it m}\,}}

\def\sub#1{_{\lower.25ex\hbox{$\scriptstyle#1$}}}
\def\sul#1{_{\kern-.1em#1}}
\def\sll#1{_{\kern-.2em#1}}
\def\sbl#1{_{\kern-.1em\lower.25ex\hbox{$\scriptstyle#1$}}}
\def\ssb#1{_{\lower.25ex\hbox{$\scriptscriptstyle#1$}}}
\def\sbb#1{_{\lower.4ex\hbox{$\scriptstyle#1$}}}

\def\gev{\,{\rm GeV}}

\def\to{\rightarrow}
\def\mh{\ifmmode m\sbl H \else $m\sbl H$\fi}
\def\mch{\ifmmode m_{H^\pm} \else $m_{H^\pm}$\fi}
\def\mt{\ifmmode m_t\else $m_t$\fi}
\def\mc{\ifmmode m_c\else $m_c$\fi}
\def\mz{\ifmmode M_Z\else $M_Z$\fi}
\def\mw{\ifmmode M_W\else $M_W$\fi}
\def\mws{\ifmmode M_W^2 \else $M_W^2$\fi}
\def\mhs{\ifmmode m_H^2 \else $m_H^2$\fi}
\def\mzs{\ifmmode M_Z^2 \else $M_Z^2$\fi}
\def\mts{\ifmmode m_t^2 \else $m_t^2$\fi}
\def\mcs{\ifmmode m_c^2 \else $m_c^2$\fi}
\def\mchs{\ifmmode m_{H^\pm}^2 \else $m_{H^\pm}^2$\fi}
\def\ztwo{\ifmmode Z_2\else $Z_2$\fi}
\def\zone{\ifmmode Z_1\else $Z_1$\fi}
\def\mtwo{\ifmmode M_2\else $M_2$\fi}
\def\mone{\ifmmode M_1\else $M_1$\fi}
\def\tb{\ifmmode \tan\beta \else $\tan\beta$\fi}
\def\xw{\ifmmode x\sub w\else $x\sub w$\fi}
\def\ch{\ifmmode H^\pm \else $H^\pm$\fi}
\def\lum{\ifmmode {\cal L}\else ${\cal L}$\fi}
\def\inpb{\ifmmode {\rm pb}^{-1}\else ${\rm pb}^{-1}$\fi}
\def\infb{\ifmmode {\rm fb}^{-1}\else ${\rm fb}^{-1}$\fi}
\def\epem{\ifmmode e^+e^-\else $e^+e^-$\fi}
\def\ppb{\ifmmode \bar pp\else $\bar pp$\fi}

\newskip\zatskip \zatskip=0pt plus0pt minus0pt
\def\matth{\mathsurround=0pt}

\def\atversim#1#2{\lower0.7ex\vbox{\baselineskip\zatskip\lineskip\zatskip
  \lineskiplimit 0pt\ialign{$\matth#1\hfil##\hfil$\crcr#2\crcr\sim\crcr}}}

\renewcommand{\thefootnote}{\fnsymbol{footnote}}

\begin{document} \begin{titlepage}
\setcounter{page}{1}
\thispagestyle{empty}
\rightline{\vbox{\halign{&#\hfil\cr
&ANL-HEP-PR-92-27\cr
&April 1992\cr}}}
\vspace{1in}
\begin{center}

{\Large\bf
Contributions to the W-Boson Anomalous Moments in the Two-Higgs-Doublet Model
at Collider Energies}\footnote{Work supported
by the U.S. Department of
Energy, Division of High Energy Physics, Contracts\newline
W-31-109-ENG-38 and W-7405-Eng-82.}
\medskip

\normalsize THOMAS G. RIZZO
\\ \smallskip
High Energy Physics Division\\Argonne National
Laboratory\\Argonne, IL 60439\\
\smallskip
and\\
\smallskip
Ames Laboratory and Department of Physics\\
Iowa State University\\ Ames , IA 50011\\

\end{center}

\begin{abstract}

We examine the one loop contributions arising in the Two-Higgs-Doublet
Model (THDM) to the W-boson anomalous magnetic dipole and
electric quadrupole form factors for both photon and Z
couplings relevant at collider energies. While the model
parameter and $q^2$-dependencies of these form factors are found to be
significant, the corresponding size of these corrections are relatively
small in comparison to unity. They are, however, found to be comparable in
magnitude to
the usual Standard Model loop corrections. Radiative corrections to the Higgs
particle masses and couplings due to heavy top-quarks are included in the
analysis.
\end{abstract}

%\vskip1.75in

%\noindent{(Talk given at the {\it Workshop on Photon Radiation from Quarks},
%Annecy, France, December 2-3, 1991.)}

\renewcommand{\thefootnote}{\arabic{footnote}} \end{titlepage}

%%%%%%%%%%%%%%%%%%%%%%%%%%%%%%%---- text

Many theories which go beyond the Standard Model (SM), such as the
Peccei-Quinn Model, Supersymmetry(SUSY), and ${\rm E}_6$ supersting-inspired
models,
 require the scalar
sector to be enlarged by the addition of an extra Higgs doublet
\cite{PQ,SUSY,E6}. Such Two-Higgs-Doublet Models (THDM) have an extremely rich
phenomenology\cite{HHG} as the particle spectrum now contains two neutral
CP-even states, h
and H, a neutral CP-odd state, A, and a conjugate pair of charged fields,
${\rm H}^\pm$.
Unfortunately, although there have been direct searches for such particles at
both $e^+e^-$\cite{LEPS} and hadron\cite{HADS} colliders as well as searches
using indirect techniques\cite{IND}, they have failed to be observed.
Of course, the searches so far conducted have only explored a small region of
the allowed parameter space in these models but the advent of hadron
supercolliders\cite{SSC,LHC} and $e^+e^-$ machines in the TeV range\cite{NLC}
will allow essentially all of the parameter space of current interest
to be probed.

As the detailed predictions of the SM begin to be examined, one way to look for
new physics is via loop-effects through possible small deviations
away from SM expectations. Using existing data, this has already been done by
a large number of authors\cite{bigref} for many possible SM extensions.
As has been suggested\cite{bigtri}, one
place to look for deviations in the future is the trilinear gauge couplings,
$\gamma$WW and ZWW, which are uniquely predicted in the SM. In the
$\gamma$WW case, with all
particles on-shell, loop corrections to the tree-level SM predictions were
considered some time ago\cite{rctri} for several different SM extensions
including the THDM case. In
this paper, we return to a discussion of the loop corrections to the
$\gamma$WW
vertex within the
context of the THDM, incorporating several refinements to fully cover the
parameter space and make a direct connection to the physical Higgs spectrum.
We then
 extend the previous analysis to the ZWW case and generalize to the situation
 where the $\gamma$ or Z can be off-shell. This is, of course, what
will be realized experimentally when the trilinear couplings are probed
in W-boson pair production at $e^+e^-$ colliders. In particular, we are
interested in the $q^2$-dependence of the anomalous magnetic dipole and
electric quadrupole moment form factors and how they differ from the
expectations of the SM. We find that for both the $\gamma$WW and ZWW cases,
and all
interesting values of $q^2$, both the model parameter and $q^2$-dependencies
of these
deviations are significant. Unfortunately, we also show that the size of
these deviations are always relatively small in comparison to unity although
they can be comparable in magnitude to the SM predictions themselves. It may,
however, be possible to probe these anomalous couplings at the few parts per
mil level at the NLC\cite{ALTER} given sufficient integrated luminosity.

To perform our calculation we label the momentum flow in the V($=\gamma$ or
Z)WW
vertex as shown in Fig.1. Although we allow V to be off-shell, we assume that
terms proportional to $Q^\lambda$ can be dropped since V is assumed to couple
to massless fermions. Both W's are assumed to be on-shell in the discussion
below.
Following the notation of Hagiwara et.al.
\cite{bigtri}, the relevant part of the coupling in momentum space is given by
\medskip
\begin {equation}
 g_{VWW}^{-1}\Gamma^{\lambda\mu\nu}_V = 2f_1^V p^\lambda
g^{\mu\nu}-{8f_2^V\over {M_W^2}} p^\lambda Q^\mu Q^\nu
+2f_3^V(Q^\nu g^{\mu\lambda}-Q^\mu g^{\nu\lambda})
\end {equation}
\medskip
when both the ${\rm W}^+$ and ${\rm W}^-$ are on-shell. At tree-level, the SM
gauge symmetries demand that $f_1^V=1$, $f_2^V=0$, and $f_3^V=2$
with $g_{\gamma WW}=-e$ and $g_{ZWW}=-gc_W$ where  $c_W=cos \theta_W$.
At one-loop, however, these
parameters are now shifted and are given instead by (here s is the square of V
invariant mass, i.e., $s=q^2$, which we will assume, for the moment, is
positive)
\medskip
\begin {eqnarray}
f_1^V & = & 1+\Delta g_1^V+{s\over 2M_W^2}\lambda_V \nonumber\\
f_2^V & = & \lambda_V \\
f_3^V & = & 2+\Delta g_1^V+\Delta\kappa_V+\lambda_V \nonumber
\end {eqnarray}
\medskip
so that the momentum space coupling can be written as
\medskip
\begin {eqnarray}
g_{VWW}^{-1}\Gamma_V^{\lambda\mu\nu} & = & (1+\Delta g_1^V+
{s\over 2M_W^2}\lambda_V)
[2p^\lambda g^{\mu\nu}+4(Q^\nu g^{\mu\lambda}-Q^\mu g^{\nu\lambda})] \\
& - & 8{\lambda_V\over M_W^2}p^\lambda Q^\mu Q^\nu +2[\Delta\kappa_V
+\lambda_V(
1-s/M_W^2)](Q^\nu g^{\mu\lambda}-Q^\mu g^{\nu\lambda}) \nonumber
\end {eqnarray}
\medskip
Note that both $\lambda_V$ and $\Delta\kappa_V$ are finite and
$q^2$-dependent,
whereas $\Delta g_1^V$ contains ultra-violet infinities corresponding to a
charge renormalization. (In the case where a massless scalar is exchanged
between the W's, an infra-red divergence can also be isolated in
$\Delta g_1^V$ which is then cancelled by a real emmission diagram.)
Instead of being constants, both $\Delta\kappa_V$ and
$\lambda_V$ are true form-factors.
As can be seen from this equation, the form of the
finite terms that we are interested in here ($\Delta\kappa_V$ and $\lambda_V$)
 are somewhat different than those
used by Couture et.al.\cite{rctri} in the case where the initial photon, as
well as the W-pair, are
on-shell. This problem is easily overcome by noting that the two
parameterizations are simply related via
\medskip
\begin {eqnarray}
(\Delta\kappa)_{Couture} & = & \Delta\kappa_V + \lambda_V(1-s/M_W^2)
\nonumber \\
(\Delta Q)_{Couture} & = & -2\lambda_V
\end {eqnarray}
\medskip
Our convention is now the standard one adopted by most authors in the
literature\cite{bigtri}.

The two classes of diagrams contributing to $\Delta\kappa_V$ and $\lambda_V$
at one loop in the THDM are shown in Fig.2. In the first class, the external
V couples to either a pair of W's, a pair of charged Goldstone bosons,  or one
of each and the two external W's are
connected by the exchange of either h or H. In diagrams of the second class,
the external V couples to ${\rm H}^{\pm}$ and the external W's are connected
by h, H, or A exchange. (Other classes of diagrams involving the ZhA and ZHA
vertices for the V=Z case are found to cancel due to Bose statistics.)
Denoting the contributions of these twos set of
diagrams by the $q^2$-dependent quantities
$I_1$ and $I_2$, we can write the entire THDM contribution to
either $\Delta\kappa_V$ or $\lambda_V$ symbolically as
\medskip
\begin {equation}
[fI_1(h)+I_2(H,H^{\pm})]s_{\beta-\alpha}^2 + [fI_1(H)+I_2(h,H^{\pm})]c_{\beta
-\alpha}^2 + I_2(A,H^{\pm}) -SM
\end {equation}
\medskip
In this equation, the aguments of $I_{1,2}$ represent the relevant particle
masses, $c_{\beta-\alpha}=cos(\beta-\alpha)$ and $s_{\beta-\alpha}=sin(\beta
-\alpha)$ which are the usual mixing
angle factors\cite{HHG}, and $f$ depends on the choice made for V:
\medskip
\begin {equation}
f = \left\{ \begin{array}{ll}
      1\,, & {V=\gamma} \\
      (1-2x_W)/2(1-x_W)\,, & {V=Z}
\end {array}
\right.
\end {equation}
\medskip
Also note that in Eq.(5) we have subtracted off the SM contribution to both
$\Delta\kappa_V$ and $\lambda_V$ since we are interested in how much these
quantities are changed when the SM is extended to the THDM case. To be
specific, we have subtracted the SM contributions assuming a SM Higgs mass of
200 GeV and taking the appropriate value of $q^2$. (Of course,
a different choice for the SM Higgs mass will only shift our results for
$\Delta\kappa_V$ and $\lambda_V$ by a pair of constants.) This
subtraction will be most
noticable in the case of $\lambda_V$ which would normally vanish in the limit
that $all$ the masses in the loop become large. Instead, we simply recover
the negative of the SM contribution in the present case.
Note that in the
limit that H, A, and ${\rm H}^\pm$ masses get very large and $s_{\beta-\alpha}$
 $\to$ 1, we will recover the SM limit provided that h is identified with
the Higgs scalar of the SM in which case the difference defined in Eq.(5) will
 vanish.

Since the THDM contains many unknown quantities, it is very difficult to
explore all of the parameter space. The above results, while being valid in any
THDM whose scalar sector conserves CP, are difficult to analyse numerically.
In order to simplify our results and to
reduce the number of free parameters we have assumed that the THDM Higgs
masses and couplings are as given by SUSY-version of the Model
with the one-loop heavy top-quark  mass corrections included\cite{HSRAD}.
Specifically, we have chosen the same input parameters as used
in the effective potential analysis of Barger et.al.
\cite{SSC} except that we have taken the top-quark mass to be 130 GeV. Our
results are not qualitatively sensitive to this particular choice of
parameters. With
this great simplification, the only remaining free parameters are the ratio
of the two
Higgs doublet
vacuum expecation values, $tan \beta =v_t/v_b$, and the mass of the CP-odd
field A, $m_A$. All other masses and mixing angle parameters are then fixed.
For numerical purposes, will we assume $M_W = 80.14\gev$ \cite{WMASS} and $M_Z
=
91.175\gev$ \cite{ZMASS} with $c_W = M_W/M_Z$ in the analysis that follows. The
functions $I_{1,2}$ can be expressed generically in terms of a small set of
integrals which reduce to those of Couture et al.\cite{rctri} in the limit that
$q^2 \to 0$. For example, in the $\lambda_V$ case, we define the integral
\medskip
\begin {equation}
J(r_q,r_p,r_0)=6a \int^{1}_{0} dt~t^3 (1-t) \int^{1}_{0} du ~u(1-u)
[A-Bu(1-u)]^{-1}
\end {equation}
\medskip
with $B \equiv r_qt^2$ and $A \equiv t^2-t(1+r_0-r_p)+r_0$. The $r's$ are
given by
\medskip
\begin {eqnarray}
r_q &=& q^2/M_W^2    \nonumber  \\
r_p &=& m_{{\rm H}^\pm}^2/M_W^2  \\
r_0 &=& m_0^2/M_W^2  \nonumber
\end {eqnarray}
\medskip
with $m_0$ being a generic mass for a neutral Higgs field. Note that
we have defined the dimensionless parameter, $a$, as
\medskip
\begin {equation}
a={g^2\over 96\pi^2}
\end {equation}
\medskip
which sets the typically small scale for the size of the loop-induced
anomalous couplings.
We now find that $I_{1,2}$ in Eq.(5) can now be simply expressed as
\medskip
\begin {eqnarray}
I_1^\lambda (h,H) &=& J(r_q,1,r_{h,H}) \nonumber \\
I_2^\lambda (h,H,A;H^\pm) &=& J(r_q,r_p,r_{h,H,A})
\end {eqnarray}
\medskip
in obvious notation. A similar set of relations can be developed for the case
of $\Delta\kappa_V$; definining the integrals
\medskip
\begin {eqnarray}
K(r_q,r_p,r_0) &=& -6a \int^{1}_{0} dt~\int^{1}_{0} du ~t(3ut-1)~
log|A-Bu(1-u)|     \nonumber \\
L(r_q,r_p,r_0) &=& 3a \int ^{1}_{0} dt~\int^{1}_{0} du ~t^2(4+2tr_0-2r_0)
[A-Bu(1-u)]^{-1}
\end {eqnarray}
\medskip
with $A$ and $B$ as given above, we find that
\medskip
\begin {eqnarray}
I_1^{\Delta \kappa}(h,H) &=& K(r_q,1,r_{h,H})-(1-r_q)J(r_q,1,r_{h,H})
\nonumber \\
I_2^{\Delta \kappa}(h,H,A;H^\pm) &=& K(r_q,r_p,r_{h,H,A})-(1-r_q)
J(r_q,r_p,r_{h,H,A})+L(r_q,r_p,r_{h,H,A})
\end {eqnarray}
\medskip
The above integrals are relatively easy to perform numerically once care is
given to potential imaginary parts. Since $\Delta\kappa_V$ and $\lambda_V$ are
themselves
generated only at the one-loop level, they contain no imaginary parts at this
order; such imaginary parts will, however, arise at the two-loop level.
(The imaginary part of the one-loop vertex function is in fact buried in the
quantity $\Delta g_1^V$ introduced above.) These results are seen to
simply reduce to
those obtained earlier assuming V=$\gamma$ and taking $q^2$ to zero.

In order to get a feeling for the
 relative
size of the THDM contributions we note that the SM with $m_t=130 \gev$ and
$M_H=200 \gev$ predicts $\Delta\kappa_\gamma=13.5a$ and $\lambda_\gamma =-0.6a$
for $q^2=0$. (Note that the predicted size of $\lambda_\gamma$ is more than
an order of magnitude smaller than that of $\Delta\kappa_\gamma$.)
As we will see, the THDM contributions are quite comparable to
these values in size. Unfortunately, since $a$ is a small quantity, $\Delta
\kappa_{\gamma,Z}$ will almost always be less than $\simeq$ 0.01 in
magnitude (with $\lambda_{\gamma,Z}$ even smaller) in the THDM case.
It is important to remember that such SM
contributions are subtracted out in what follows as purely SM radiative
corrections are already well understood in processes such as $e^+e^-\to
W^+W^-$ in which our anomalous couplings might be measured.

Since the $q^2=0$ situation has already been discussed previously for the
case $V=
\gamma$ \cite{rctri} we will mention it here only to illustrate the $q^2$
dependence of the two form factors. It will, however, show many features in
common with cases where $q^2 \neq 0$ and where V=Z. Figs.3a and 3b show
$\Delta\kappa_\gamma$ and $\lambda_\gamma$ as functions of $m_A$ for different
choices of $tan \beta$ assuming $q^2$=0. The first thing we observe
is the apparent lack of
sensitivity to the choice of $tan \beta$, a property that will persist in
all other
cases as we will see below. Both $\Delta\kappa_\gamma$ and $\lambda_\gamma$
have their greatest magnitudes when $m_A$ is small. The reason for this is
clear; as $m_A$ increases the corresponding masses for H and ${\rm H}^\pm$ also
become large leaving only h light. The h contribution is then largely cancelled
by the usual SM subtraction; this is particularly true in
the $\lambda_\gamma$ case
where decoupling in the heavy mass limit is most effective. We also note
again the
general feature that the relative sizes of $\Delta\kappa_\gamma$ and
$\lambda_\gamma$ generally differ by more than an order of magnitude; this
effect will occur in all the cases we will examine below. As advertised, the
THDM contributions to both of these parameters are generally comparable to
their SM values in magnitude, particularly for small values of $m_A$. These
results were found to be only moderately sensitive to the existence of heavy
top-quark loop corrections to the Higgs' masses and couplings, particularly,
at smaller values of $m_A$, and only midly sensitive to the particular SUSY
parameter values used to calculate these corrections. These properties
continue to valid in all of the results we present below.

To verify the lack of $tan \beta$ sensitivity, we show in Figs.3c and 3d the
values of $\Delta\kappa_\gamma$ and $\lambda_\gamma$ as functions of $tan
\beta$
for different values of $m_A$ assuming $q^2$=0. Quite generally, we see
that the values of
these parameters are
approximately constant for $tan \beta$ larger than unity which is just the
region in which we expect the SUSY THDM mass and coupling realtions
from the effective potential approach that we
have used in these calculations to be most valid \cite{SSC}.

What happens when we increase $q^2$ and we are no longer probing the static
moments? Figs.4a-d and 5a-d show the results akin to Figs.3a-b for different
values of $q^2$ and we see a significant change in the shape of the curves when
the W-pair threshold is exceeded. The general lack of sensitivity to
$tan \beta$ is still evident except for some very small ranges of $m_A$; this
we have examined explicitly for several fixed values of $q^2$.  In the
case of $\lambda_\gamma$, we note that the magnitude decreases significantly as
$q^2$ increases.
This behaviour might have been anticipated by the results in Eq.(10).
$\Delta\kappa_\gamma$ also decreases in magnitude once the
value of $q^2$ exceeds the W-pair threshold but does not decrease significantly
beyond $\sqrt s$=200 GeV. To explore this behaviour further,
we display the explicit
$q^2$-dependencies of both $\Delta\kappa_\gamma$ and $\lambda_\gamma$ in Figs.
6a-b assuming $tan \beta$ =2 for different values of $m_A$. (Due to the general
 overall
lack of sensitivity to $tan \beta$, these results are largely independent of
the value of
this parameter.)  Except for the region near the W-pair threshold,
these functions are
generally smooth and $\lambda_\gamma$ is seen to tend to zero in the large
$q^2$ limit as expected. The asymptotic behavior of $\Delta\kappa_\gamma$ is
somewhat
more complex in this limit. We do, however, see the general result that the
$q^2$-dependence of both $\Delta\kappa_\gamma$ and $\lambda_\gamma$ in the
THDM cannot be
ignored in that the deviations expected for the $static$ moments in the THDM
are quite different from those one would measure at a high energy
collider producing W-pairs. Unfortunately, although the size of these
corrections we have found are
comparable to the SM results themselves, they are numerically quite small.

Do these general features persist when we look at the case V=Z as we might
naturally expect? Fig.7 answers this question in the affirmative; here we see
repeated lack of dependence of our results on $tan \beta$ and the drastic
reduction in magnitude of $\lambda_Z$ as $m_A$ increases. Generally, the
magnitudes we find for $\lambda_Z$ and $\Delta\kappa_Z$ are comparable to
what was seen in the case V=$\gamma$. In fact, the numerical results for the
two
quantities in the V=Z and V=$\gamma$ cases are found to be very similar for
most values of the parameters.  Fig.8 shows that the $q^2$-dependence of the
form factors in both cases is also extremely similar as one might expect.
Thus in the case of W-pair production at $\sqrt s$=200 GeV, the anticipated
 size of $\lambda_\gamma$ amd $\lambda_Z$ (as well as $\Delta\kappa_\gamma$
 and $\Delta\kappa_Z$) are quite comparable in both the SM and the THDM.

We have examined the contributions to the anomalous WWZ and WW$\gamma$
trilinear couplings within the context of the THDM over a wide range of
model parameters and values of $q^2$. We have incorporated the one-loop
radiative corrections due to a heavy top-quark into the mass and
coupling relationships amongst the various physical Higgs fields.
The major results that we have obtained in our analysis are as follows:

(${\it i}$) Both $\Delta\kappa_V$ and $\lambda_V$ were found to be relatively
insensitive to the choice of $tan \beta$ once this quantity was greater than
unity. The sensitivity to the value of $m_A$ was particularly stong
especially for $m_A$ values below about 200 GeV. The usual SM results were
 obtained in the limit of large $m_A$. Although the THDM contributions did not
lead to order of mamnitude changes in the anomalous moment for factors, we did
see that the additional contributions could be quite comparable to the SM
results.

(${\it ii}$) The anomalous couplings were seen to be of comparable magnitude
in both the V=Z and V=$\gamma$ cases. Thus it would be impossible to seperate
these two sources of anomalous couplings using $e^+e^-$ data alone. Although
the values we have obtained for the anomalous moment form factors in the THDM
remain reasonably small, measurements taken at the NLC may help to constrain
the THDM parameter space.

(${\it iii}$) The two form factors we examined were observed to exhibit a
strong
$q^2$ dependence. Their values, as would be measured via W-pair production at
an $e^+e^-$ collider, are substantially different than their static values
obtained at $q^2$=0. $\lambda_V$ was found to decrease in magnitude quite
rapidly with increasing $q^2$.

(${\it iv}$) The influence of incorporating the heavy top-quark loop
corrections to the Higgs' masses and couplings was found to be significant
particularly in the range of small $m_A$ values. The specific results were,
however,
quite insensitive to the particular choice of the SUSY parameters that were
employed.

Precision measurements of the anomalous moment couplings of the W at $e^+e^-$
colliders and elsewhere may lead to the first indications of the existence of
new physics beyond the Standard Model.

\newpage
\vskip.25in
\centerline{ACKNOWLEDGEMENTS}

The author would like to thank JoAnne L. Hewett for extensive
discussions related to this work as well as John Ng and Gilles Couture
for getting him interested in this subject and
for past collaborative efforts.
This research has been supported by the U.S. Department of Energy,
Division of High Energy Physics, Contracts W-31-109-38 and W-7405-Eng-82.

\newpage

%
%%%%%%%%%%%%%%%%%%--- References
%%%%%%%%%%%%%%%%%%%%%%%%%%%%%%%%%%%%%%%%%%%%%%%%%%%%%%%
\def\MPL #1 #2 #3 {Mod.~Phys.~Lett.~{\bf#1},\ #2 (#3)}
\def\NPB #1 #2 #3 {Nucl.~Phys.~{\bf#1},\ #2 (#3)}
\def\PLB #1 #2 #3 {Phys.~Lett.~{\bf#1},\ #2 (#3)}
\def\PR #1 #2 #3 {Phys.~Rep.~{\bf#1},\ #2 (#3)}
\def\PRD #1 #2 #3 {Phys.~Rev.~{\bf#1},\ #2 (#3)}
\def\PRL #1 #2 #3 {Phys.~Rev.~Lett.~{\bf#1},\ #2 (#3)}
\def\RMP #1 #2 #3 {Rev.~Mod.~Phys.~{\bf#1},\ #2 (#3)}
\def\ZP #1 #2 #3 {Z.~Phys.~{\bf#1},\ #2 (#3)}
\def\IJMP #1 #2 #3 {Int.~J.~Mod.~Phys.~{\bf#1},\ #2 (#3)}

\newpage

%%%%%%%%%%%%%%%%%%%%%%%--- figures
%
{\large\bf Figure Captions}
\begin{itemize}

\item[Figure 1.]{Feynman diagram for the VWW vertex with the momenta labeled
as used in the text.}
\item[Figure 2.]{The two classes of triangle diagrams responsible
for generating the anomalous moments of the W-boson in the THDM.}
\item[Figure 3.]{The $\gamma$WW anomalous moments for the case $q^2$=0.
(a) $\Delta\kappa_{\gamma}$
as a function of $m_A$ for $tan \beta$ =1, 2, 5, 10,and 30; (b)
same as (a) but for $\lambda_\gamma$; (c) same as (a) but as a function of
$tan \beta$ for $m_A$=50 (top solid), 100(dashed-dot), 250(dashes), 500(
dots), or 1000(bottom solid)$\gev$; (d) same as (c) but for $\lambda_\gamma$
with the top solid curve now corresponding to $m_A$=1000$\gev$ and the order
of the subsequent curves in (c) now reversed.}
\item[Figure 4.]{$m_A$ dependence of $\Delta\kappa_\gamma$ for $\sqrt s$=
(a) 100; (b)200; (c)500; or (d)1000$\gev$ with $tan \beta$ as given in Fig.3.}
\item[Figure 5.]{Same as Fig.4 but for $\lambda_\gamma$.}
\item[Figure 6.]{$\sqrt s$ dependence of (a) $\Delta\kappa_\gamma$ and (b)
$\lambda_\gamma$ assuming $tan \beta$=2. In (a), from top to bottom on the
left hand side, the curves correspond to $m_A$=50, 100, 200, 500, or 1000
$\gev$ while in (b) the same order is reversed.}
\item[Figure 7.]{$m_A$ dependence of $\Delta\kappa_Z$ and $\lambda_Z$ for the
same values of $tan \beta$ as in Fig.3 with either $\sqrt s$
=200, 500, or 1000 $\gev$.}
\item[Figure 8.]{Same as Fig.6 but for $\Delta\kappa_Z$ and $\lambda_Z$.}

\end{itemize}

\end{document}